\documentclass{article}

\begin{document}

\title{Fun in 2+1}

\author{S. Deser\\
\emph{Walter Burke Institute for Theoretical Physics,}\\
\emph{California Institute of Technology}\\
\emph{Pasadena, CA 91125, USA}\\
\emph{Physics Department, Brandeis University}\\
\emph{Waltham, MA 02454, USA}\\
\emph{deser@brandeis.edu}}
\maketitle


\begin{abstract}
\noindent To celebrate Roman Jackiw’s eightieth birthday, herewith some comments on gravity and gauge theory models in $D=3$, the chief focus of our many joint efforts.  
\end{abstract}




\section{Introduction}
It is a pleasure to dedicate this note, on this round birthday, to my long-time friend and collaborator Roman Jackiw. We were born eight years apart at about the same latitude, if slightly different longitudes, in prewar Poland, but first met when he arrived in Cambridge as a Harvard Junior Fellow, after a Cornell PhD --- and a co-authored graduate textbook --- with Hans Bethe. That guaranteed his technical skill and mastery of theory, all of which have been amply borne out by his subsequent contributions. Roman has been an ornament to MIT --- and the Boston area physics community --- ever since. Our own collaboration spanned a quarter century, with fifteen papers garnering some 6500 citations to date. Even so, our joint best seller is only his \#2 work (\#1 would be the ABJ chiral anomaly). These bare statistics miss all the fun we had, ranging from lecturing at a Brazilian school in far-off Ouro Prieto to soirees in his and So-Young Pi’s Beacon Hill apartment, to our many violent discussions while collaborating and to the altogether different tenor of the interactions during our related efforts with Gerard 't Hooft. 

Our work over those decades had an unplanned, but retrospectively manifest, theme: Quantum field theories and gravity models in lower --- especially $D=3$ --- dimensions. It proved extremely fruitful, giving rise to numerous novel consequences, including massive yet gauge invariant models \cite{deser82a,deser82b} --- the first new classical mass generating mechanism --- amongst all the other new Chern--Simons (CS) physical and mathematical effects, and separately, what we called the dynamics of flat space, finding life in --- seemingly empty --- GR in $D=3$ \cite{deser84a}. These topics having continued to snowball in popularity, this may be a good point to comment on some aspects of their evolution. 

\section{Why $D=3$?}
As a child, I recall seeing a popular (prewar) book entitled ``Why split atoms?", one that I assumed pled for less cruelty to them. Here I ask ``why work in $D=3$?" in a rather more positive vein. The obvious Everest defense is of course ``because it’s (sometimes effectively) there" but there are good reasons aside from the fun and learning we derive in seeming detours from reality. For example, as I write this, a new posting suggests the use of the strong fields in various gravitational collapses to look for the, slightly generalized $D=4$ extension of, CS gravity whose antecedents can also be traced to ($D=10$!) string theory. The totally different CS application to the quantum Hall effect is instead about planar condensed matter physics. Then there is the enormous and still growing, mathematical physics, pure CS industry, let alone its supersymmetric versions whose existence are a model’s sine qua non these days, and brings the whole CS notion to fermions as well \cite{deser83}! Our revival of $D=3$ GR had another spinoff: the BTZ industry --- sourceless black holes entirely due to identifying points in pure AdS \cite{banad92}. Just as $D>4$ physics has been a great source of $D=4$ insights, so clearly is $D=3$.

\section{$D=3$}
In $D=4$, Einstein gravity (GR) is itself a dynamical system whose excitations must have positive energy for stability; this is guaranteed (only) with the canonical gravitational sign choice. As a far from trivial dividend, this same choice leads to attraction between like masses \cite{deser05}. But $D=3$ Einstein gravity \cite{deser84a} is neither dynamical nor does it mediate forces between masses; the former because Riemann and Einstein tensors are equivalent here (they are double duals of each other), so empty ($G_{mn}=0$) space is flat; the latter because space’s flatness between sources means it cannot transmit any local (but only global) interactions between them. Nevertheless, there remains a rich global source interaction, including their quantum scattering \cite{deser88,hooft88}! The cosmological, (A)dS, extensions \cite{deser84b} of $D=3$ GR are correspondingly different from their $D=4$ counterparts as well. 

Separately, our totally different, odd-dimensional, $D=3$ extensions of gauge theory and of GR, introduced a physical role for their respective first and third derivative order Chern--Simons terms, and incidentally forcing the sign of the (now lower derivative) Einstein component to be opposite to that in $D=4$. These models have deep ``topological" roots: their gauge and gravity terms are the 3D ``descents" $K^4$ of the 4D axial invariant densities $F^*F$ and $R^*R$, each of which is a total divergence $\partial_\mu K^\mu$.

\section{3 topics in 3D}
So much for generalities; let us now briefly consider some specifics, starting with aspects of 3D GR AdS solutions and their relations to the very different nature of the (subsequent) BTZ solutions. Consider the frame in which the source-free AdS metric is
\begin{equation}
ds^2= -F dt^2 + F^{-1} dr^2 + r^2 d\theta^2,   \quad  F=1+ \left(\frac{r}{l}\right)^2.
\end{equation}
In Ref.~\cite{deser84b}, where $\Lambda= l^{-2}=0$, it was shown that a (positive mass $m$) particle action produces the value 
\begin{equation}\label{eq2}
F= 1- 4Gm + 0 \quad\rightarrow\quad  F = 1-4Gm + \left(\frac{r}{l}\right)^2.
\end{equation}
 Specifically, (\ref{eq2}) first shows the $\Lambda=0$ point source solution, then how it generalizes to AdS for a positive mass point source and gravitational constant $G$. The conventional (``$D=4$") sign is $G>0$. [We actually generalized to more sources, but that is irrelevant here.] Instead, the BTZ solution is given in the same frame by
\begin{equation}
F^* = -M + \left(\frac{r}{l}\right)^2,
\end{equation}
where $M$ is a positive constant. It is this negative sign choice that provides the black hole (BH) interpretation, a bit like that of the (``1/2") dS intrinsic horizon, where instead the $r^2$ term turns negative and its ``$F$" $=1- (r/l)^2$. The BTZ solution has no explicit matter source; rather, it is pure AdS  but with certain points identified. [In $D=4$, the BH sign is automatically related to overall (necessarily positive) mass, this coefficient of $1/r$ being the total energy of the system, with, or even without, physical matter sources.]  Let us instead consider this metric In terms of our physical point sources as in (\ref{eq2}). The combination $4Gm$ must exceed unity for negative $-M$ (note that only the product $Gm$ enters, BUT $m$ is necessarily $>0$, so only the conventional positive sign of $G$ has a chance.) Recall also that there is no gravitational contribution to the total mass in $D=3$, as noted earlier. As already shown in Ref.~\cite{deser84a}, these are forbidden unphysical, values. Indeed, here is the relevant explicit quote: ``In the above we have taken $m < 1/4G$. When $m$ exceeds this limit, the metric near the particle becomes singular, e.g., the distance from the particle to any other point diverges as $r^{1-4Gm}$. This limitation on the mass may also be understood in terms of the formal equivalence between $a  \leftrightarrow -a$ and the coordinate inversion $r \leftrightarrow l/r$ implied by (2.8). In other words, a high-mass particle at the origin with $m > 1/4G$ is actually a particle at infinity with acceptable mass $1/2G - m$. The geometrical counterpart of this argument is given in Section V, where it is also shown that composite (many-body) sources can have a higher (up to $1/2G$) total mass."  Nothing changes in this analysis for AdS gravity; for example the diverging distance is a purely local effect. The special value $4Gm=1$ is also analyzed there to be a strange cylinder. We conclude that the BTZ solution, while mathematically correct as a pure AdS space with those identifications, cannot be constructed from physical, positive mass, sources. The authors indeed state that theirs are solutions of the source-free Einstein equations, i.e., without matter sources, and specifically without the cone structure emanating from point matter. We have in fact speculated that these solutions are separated from their normally sourced counterparts by a ``superselection" rule \cite{deser10}.

Next, a brief look at closed timeline curves (CTC), a.k.a. time travel in pure GR. This subject, first worried about by Einstein, even before completing GR, as a possible disease of dynamical, as against fixed, spacetime, was of course made concrete by Godel’s contribution to Einstein’s 70th birthday(!), but was not as scary as local CTC discovered later. 

In Refs.~\cite{deser92a} and \cite{deser92b}, 't Hooft, Roman and I responded to a purported Godelian $D=3$ solution involving string sources \cite{gott91}. True to Einstein’s original intuition that physically admissible sources would not produce CTC, we showed that the threat was indeed unphysical here as well. Ironically, many years later, the local CTC menace was to arise in the more dangerous $D=4$ massive gravity context \cite{deser13a,deser13b}, this time showing it to be unphysical as well, a welcome echo of our 3D work!

As a final topic in our sampler, we describe the remarkably simple --- in first order form, like its SUGRA predecessor --- locally supersymmetric TMG (its gauge field counterpart is of course simpler still) \cite{deser83}. There are obviously separate fermionic companions to the Einstein and CS parts: the former is, also obviously, the massless Rarita--Schwinger action in a gravitational background with \emph{a priori} independent dreibein and connections to allow for torsion as in $D=4$, and equally obviously as devoid of excitations as GR. Instead, the friend of the CS term must be of second derivative order (always one less than its bosonic equivalent), and all fermionic terms must depend only on the (dual) field strength $f^\mu = \epsilon^{\mu\nu\lambda} D_\nu \psi_\lambda$ (the covariant derivative in the curl is thus only with respect to the spinorial index, namely a connection term) fo retain the original local gauge invariance under $\Delta\psi_\mu=\partial_\mu a(x)$, where $a(x)$ is the fermionic gauge parameter. This, together with the requirement that it must share the fermonic equivalent of the conformal-invariance of the CS term, forces it to be 
\begin{equation}
L^{CS}_{3/2} \sim m^{-1} \int d^3x f^\mu \gamma_\nu  \gamma_\mu f^\nu;
\end{equation}
the overall coefficient likewise shares its inverse length dimension. Indeed, even the term’s descent from a $D=4 \partial_\mu K^\mu$ is nicely reproduced. All this is gratifying, since there was no \emph{a priori} guarantee that a super version would exist, particularly one so nicely matching the topological bosonic part’s properties. 

\section{Envoi}
Many more happy years to you and to lower dimension physics progress!

\section*{Acknowledgements}
This work was supported by the U.S. Department of Energy, Office of Science, Office of High Energy Physics, under Award Number de-sc0011632.


\end{document}